\documentclass[12pt]{article}
\newcommand{\Comment}[1]{{}}
\usepackage[textwidth = 450 pt, textheight = 630 pt]{geometry}
\usepackage[parfill]{parskip}
\usepackage{amssymb,euscript,amsmath,amsfonts}
\usepackage{tikz,tikz-cd,url}
\usetikzlibrary{decorations.pathreplacing,calligraphy}
\usepackage{xcolor}
\usepackage{graphicx}
\usepackage{color}
\usepackage{bbm}
\usepackage{tensor}
\usepackage{slashed}
\usepackage{bigints}
\usepackage{mathrsfs}

\newcommand{\bb}[1]{\mathbb{#1}}

\renewcommand{\cal}[1]{\mathcal{#1}}

\newcommand{\inv}[1]{\frac{1}{#1}}

\newcommand{\brac}[1]{\left(#1\right)}

\newcommand{\mn}{\mu\nu}

\newcommand{\bbm}[1]{\mathbbm{#1}}

\allowdisplaybreaks

\definecolor{MyDarkBlue}{rgb}{0.15,0.15,0.45}
\usepackage[linktocpage=true]{hyperref}
\hypersetup{
colorlinks=true,
citecolor=MyDarkBlue,
linkcolor=MyDarkBlue,
urlcolor=MyDarkBlue,
pdfauthor={Neil Lambert },
pdftitle={ },
pdfsubject={hep-th}
}

\usepackage{hypernat}
\usepackage{physics}
\usepackage{accents}
\usepackage{bm}
\usepackage[sorting=none, style=numeric-comp]{biblatex}
\begin{document}

 \centerline{\Huge  {The M5-Brane Limit of}}
 \vskip 12pt
 \centerline{\Huge {Eleven-Dimensional Supergravity}}
 \vskip 12pt

\centerline{\LARGE \bf {\sc  }} \vspace{2truecm} \thispagestyle{empty} \centerline{
    {\large {{\sc Eric~Bergshoeff${}^{\,a \,}$}}}\footnote{E-mail address: \href{mailto:e.a.bergshoeff@rug.nl}{\tt e.a.bergshoeff@rug.nl} },
    {\large {{\sc Neil~Lambert${}^{\,b \, }$}}}\footnote{E-mail address: \href{mailto:neil.lambert@kcl.ac.uk}{\tt neil.lambert@kcl.ac.uk} } and  {\large {{\sc Joseph~Smith${}^{\,b \, }$}}}\footnote{E-mail address: \href{mailto:joseph.m.smith@kcl.ac.uk}{\tt joseph.m.smith@kcl.ac.uk} }}

\vspace{1cm}
\centerline{${}^a${\it Van Swinderen Institute}}
\centerline{{\it University of Groningen}}
\centerline{{\it Nijenborgh 3}}
\centerline{{\it 9747 AG Groningen, The Netherlands}}

\vspace{1cm}
\centerline{${}^b${\it Department of Mathematics}}
\centerline{{\it King's College London }} 
\centerline{{\it The Strand }} 
\centerline{{\it  WC2R 2LS, UK}} 

\vspace{1.0truecm}

\thispagestyle{empty}

\centerline{\sc Abstract}
\vspace{0.4truecm}
\begin{center}
\begin{minipage}[c]{360pt}{
    \noindent}

 We construct the M5-brane limit of eleven-dimensional supergravity.  The resulting action is invariant under Galilean boosts and has a local scale symmetry. We also consider the limit of the equations of motion where we   recover a Poisson-like equation arising from an M5-brane source but which does not follow from the non-relativistic action. We argue that the resulting theory describes gravitational fluctuations around a stack of M5-branes, represented by a trivial Minkowskian spacetime, but where the number of M5-branes is determined by the flux of a Lagrange multiplier field.

\end{minipage}
\end{center}

\newpage
\tableofcontents

\section{Introduction} \label{sect: introduction}

Eleven-dimensional supergravity, first constructed in \cite{Cremmer:1978km}, plays a central role in our understanding of M-theory. Originally constructed as a classical action in the maximal dimension where a supersymmetric theory (with spins less than or equal to two) is possible \cite{Nahm:1977tg}, it was later realised that it provides the low energy effective action of a complete quantum theory of gravity   whose natural dynamical objects are M2-branes and M5-branes \cite{Townsend:1995kk,Hull:1994ys,Witten:1995ex}. In particular the M2-brane arises by including a source that couples  to the dynamical 3-form $C_3$ \cite{Bergshoeff:1987cm} whereas the M5-brane appears as a solitonic object coupling magnetically to $C_3$ \cite{Gueven:1992hh}.

Recently, considerable insight has been made by considering certain non-relativistic limits associated to $p$-branes in string and M-theory  {generalizing the original work \cite{Gomis:2000bd,Danielsson:2000gi}}. In such a limit spacetime is split in  $(p+1)$ longitudinal dimensions and the remaining transverse directions. The longitudinal directions are scaled by a dimensionless  factor $c\to\infty$ whereas the transverse dimensions are scaled by a  non-positive power of $c$, determined by the nature of the limit (string, D-brane or M-brane).  
After such a limit  the resulting geometry is no longer Riemannian but rather a so-called  {p-brane} Newton-Cartan geometry \cite{Andringa:2012uz,Bergshoeff:2023rkk}. In these geometries only the spacetime along the longitudinal directions is Lorentzian  and their coordinates cannot be transformed into transverse directions, whereas the   transverse coordinates admit rotations and boost-like transformations involving the longitudinal coordinates. From the dynamical point of view taking Newton-Cartan limits in gravitational theories based on an Einstein-Hilbert action leads to novel non-Lorentzian gravitational theories.\,\footnote{{For a review on non-relativistic gravity and supergravity, see \cite{Hartong:2022lsy} and \cite{Bergshoeff:2022iyb}, respectively.}}  {Examples of such limits are the string limit of ten-dimensional $\mathcal{N}=1$ supergravity} \cite{Bergshoeff:2021tfn}, the string limit of type IIB supergravity \cite{Bergshoeff:2023ogz} and the M2-brane limit of eleven-dimensional supergravity  
\cite{Blair:2021waq,Bergshoeff:2023igy,Bergshoeff:2024nin}.  The main aim of this paper is to construct the M5-brane limit of eleven-dimensional supergravity. {This is the first example where it is explicitly shown how the same supergravity theory allows two different p-brane   limits where each limit describes the geometry close to  a stack of 2-branes or 5-branes, respectively.} 

One can also consider the effect of these limits in the gauge theories that live on multiple branes. Typically these limits lead to dynamical systems where the motion is restricted to take place on the moduli space of a particular BPS configuration \cite{Lambert:2024uue,Lambert:2024yjk}.  Physically these correspond to looking at small, slow, dynamics around a heavy soliton. Thus the $p$-brane Newton-Cartan limit is really a BPS limit, capturing the dynamics of a Lorentzian theory around a BPS soliton \cite{Avila:2023aey,Blair:2023noj,Blair:2024aqz,Harmark:2025ikv,Guijosa:2025mwh}.    
Furthermore, it is hoped that by taking matching $p$-brane limits of string/M-theory and brane worldvolume theories one will find new, possibly simpler, examples of gravity/gauge dualities with more general geometries \cite{Fontanella:2024kyl,Fontanella:2024rvn,Lambert:2024uue,Lambert:2024yjk,Lambert:2024ncn}. 

 {Taking a non-relativistic limit,} the action is expanded in powers of $c$. Terms with positive powers are diverging and must be removed or cancelled. Once this is done one can take the limit  $c\to\infty$ so that only  the $c^0$ terms survive. In this way one finds a theory with a natural Liftshitz-like rigid scale symmetry $c\to \lambda c$ for any positive constant $\lambda$. However in the case of supergravity theories a common feature   is that this rigid scale symmetry is enhanced to a {\sl local} symmetry, where $\lambda$ can be an arbitrary  function of spacetime \cite{Bergshoeff:2021bmc,Bergshoeff:2024ipq}. Such a  local scale invariance of the action leads to a redundancy in the resulting equations of motion. As a consequence  one finds that the dynamical system is not fully determined by the equations of motion arising from the  {non-relativistic} action alone. However by examining the limit of the original equations of motion one can  identify  a Poisson-like equation which replaces the missing equation of motion \cite{Bergshoeff:2021bmc,Bergshoeff:2024ipq}.

The origin of this local scale symmetry is somewhat mysterious and we will see it  occur again here for the M5-brane limit. Nevertheless it plays an important role when matching the symmetries of the supergravity to those of the field theory. In particular using the local scale symmetry one can find the same symmetries that are normally associated with $AdS$ duals without having $AdS$ factors in the geometry \cite{Lambert:2024ncn}, which opens up the possibility of new arenas for gauge/gravity duality.

The rest of this paper is organised as follows. In section \ref{sect: M2-brane limit review} we will review eleven-dimensional supergravity and the M2-brane limit of \cite{Blair:2021waq,Bergshoeff:2023igy,Bergshoeff:2024nin}, emphasizing various features, some of which are shadowed in the M5-brane case. In section \ref{sect: M5-brane limit of action} we present the M5-brane limit of the eleven-dimensional supergravity including a discussion of the resulting symmetries which again include a local scale symmetry. Section \ref{sect: equations of motion limit} revisits the M5-brane limit at the level of the equations of motion.   {In particular, we will show how to obtain the Poisson equation, which does not follow from the non-relativistic action, by considering sub-leading terms in the expansion of the equations of motion.}   
In section \ref{sect: properties of the limit} we will discuss some physical features of the action and in section \ref{sect: conclusion} we give our conclusions. We also include an appendix with details of the non-relativistic expansion of the curvature.

\section{Eleven-Dimensional Supergravity and the M2-Brane Limit} \label{sect: M2-brane limit review}

In this section we will briefly review (the Bosonic part of) eleven-dimensional supergravity and its M2-brane limit  emphasizing those aspects that are relevant for the discussion of  the M5-brane limit in the next section. Our starting point is the following relativistic action for the bosonic supergravity fields $g_{\mu\nu}$ and $C_3$\,:
\begin{align}\label{11dsugraaction}
    S_{11d} = \inv{2 \kappa_{11}^2} \bigintssss d^{11}x \sqrt{-g} R(g) - \inv{4 \kappa_{11}^2} \bigintssss \brac{ F_4 \wedge *_g F_4 + \inv{3} C_3 \wedge F_4 \wedge F_4} \ .
\end{align}
Here, $\kappa_{11}$ is the gravitational coupling constant, $R(g)$ is the Ricci scalar and  $F_4 = dC_3$. The relativistic equations of motion that follow from this action are given by
\begin{subequations}
\begin{align}
    R_{\mn} - \inv{2} R \, g_{\mn} &= \Theta_{\mu\nu}
    \ , \\ \label{eq: 3-form relativistic equation of motion}
     d *_g F_4 + \inv{2} F_4 \wedge F_4 &= 0  \ ,
\end{align}
\end{subequations}
with the energy-momentum tensor $\Theta_{\mu\nu} $ given by
\begin{subequations}
\begin{align}
    \Theta_{\mn} &= \theta_{\mn} - \inv{8} g_{\mn} g^{\rho\sigma} \theta_{\rho\sigma}\,,   \\
    \theta_{\mn} &= \inv{12} g^{\rho_2 \sigma_2} g^{\rho_3 \sigma_3} g^{\rho_4 \sigma_4} F_{\mu\rho_2 \rho_3 \rho_4} F_{\nu \sigma_2 \sigma_3 \sigma_4} \ .
\end{align}
\end{subequations}
On-shell, we can define a six-form potential $C_6$ by the duality relation
\begin{equation} \label{eq: relativistic duality relation}
    G_7 \equiv d C_6 - \inv{2} C_3 \wedge F_4 = *_g F_4 \ .
\end{equation}
Note that, due to the presence of the Chern-Simons term inside the curvature $G_7$, we cannot write the eleven-dimensional supergravity action in terms of the six-form potential $C_6$ only.

To define the M2-brane limit \cite{Blair:2021waq,Bergshoeff:2024nin} we first decompose the eleven-dimensional flat index $\hat A$ into 3 longitudinal and 8 transverse directions according to $\hat A = (A,a)$ with $A=(0,1,2)$ and $a = (3, \dots ,11)$. The indices $(A,a)$ realize a manifest SO(1,2) $\times$ SO(8) symmetry. Introducing a Vielbein field $E_\mu{}^{\hat A}$ defined by
\begin{equation}
g_{\mu\nu} = E_\mu{}^{\hat A } E_\nu{}^{\hat B} \eta_{\hat A\hat B}\,,
\end{equation}
we make  the following redefinitions of this Vielbein field and the 3-form gauge potential $C_3$:\,\footnote{In the conventions of \cite{Bergshoeff:2024ipq}, we take $\alpha=1\,, \beta =-1/2, \gamma=0$ and $\xi = 3$.}
\begin{subequations} \label{eq: M2-brane limit field redefinitions}
\begin{align}
    E^{\hat{A}}_{\mu} &= \brac{c \tau^A_{\mu}, c^{-1/2} e^a_{\mu}} \ ,\\
    C_3 &= \frac{c^3}{3!} \epsilon_{A_1 A_2 A_3} \tau^{A_1} \wedge \tau^{A_2}\wedge \tau^{A_3} + c_3 \ ,
\end{align}
\end{subequations}
where $c$ is a dimensionless contraction parameter. The fields $\tau_{\mu}^A$ and $e_{\mu}^a$ are called the longitudinal and transverse Vielbeine, respectively.

Substituting the redefinitions \eqref{eq: M2-brane limit field redefinitions} into  the action \eqref{11dsugraaction} leads to an expansion of the form
\begin{equation}
S_{11d} = c^3 S_3 + c^0 S_0 + c^{-3} S_{-3} + \dots \,.
\end{equation}
We see that there is a potential divergent term proportional to $c^3$. The Einstein-Hilbert term contributes to this  $c^3$ term as follows:
\begin{subequations}
\begin{align}
    \sqrt{-g} R &= c^3 \, \Omega R^{(3)} + \Omega R^{(0)} + O(c^{-3}) \ , \\
    R^{(3)} &= - \inv{4} \eta_{AB} T_{ab}^A T^B_{ab} \ ,
\end{align}
\end{subequations}
where $\Omega\equiv \det (\tau^A,e^a)$  is the non-relativistic measure and $T^A$ is defined by
\begin{equation}
T^A = d\tau^A  \,.
\end{equation}
The tensor $T^A$ is not covariant under longitudinal Lorentz rotations but the special projection $T_{ab}^A$ defined by
\begin{equation}
T_{ab}^A \equiv e_a{}^\mu e_b{}^\nu T_{\mu\nu}^A\,.
\end{equation}
is covariant. These special projections are called intrinsic torsion tensor components. 

We now consider how the kinetic term of $C_3$ contributes to the $c^3$ term. Using the redefinition of $C_3$ given in \eqref{eq: M2-brane limit field redefinitions} we write
\begin{equation}
F_4 = c^3 F_4^{(3)} + F_4^{(0)}\,.
\end{equation}
We use here a notation where the upper-index refers to the power of $c$ in the curved index basis and where flat indices are obtained by using the longitudinal and transverse Vielbeine. Using this notation,  we find that there are two terms that contribute to the $c^3$ term: the square of $F^{(3)}_{ABab} = T_{ab}^C\epsilon_{ABC}$ and the square of $F^{(0)}_{abcd} = f_{abcd}$ with $f_4 = dc_3$.\,\footnote{Similar to the intrinsic torsion components $T_{ab}^A$, the transverse components $f_{abcd}$ are covariant under boost transformations whereas the general components $f_4=dc_3$ are not.} The first term cancels against the contribution of the Einstein-Hilbert term to the $c^3$ term. We are then left with the square of the second term. This is particular for the membrane in eleven dimensions. For instance, for strings in ten dimensions a similar term arising  from the Kalb-Ramond kinetic term is sub-leading.

We are not yet done. Also the Chern-Simons term in the action \eqref{11dsugraaction} contributes to another $c^3$ term quadratic in $f_{abcd}$. It is a different term than the one coming from the $C_3$ kinetic term and contains an SO(8) Levi-Civita tensor.  The two terms together combine into a single term that is quadratic in the anti-selfdual field-strength $f^{(-)}$ as follows:
\begin{equation}\label{divergence1}
S_3  = - \inv{4!} \int d^{11} x \,  \Omega\,   f^{(-)}{}^{abcd}  f^{(-)}{}_{abcd} \,.\\
\end{equation}
This kind of divergence can be controlled by applying a Hubbard-Stratonovich transformation
which leads to an equivalent action $S^\prime$, containing an anti-selfdual auxiliary field $\lambda_{abcd}$\,,  whose expansion in powers of $c^3$ is:
\begin{subequations}
\begin{align}
&S^\prime_3  = 0 \,,\\
&S^\prime_0  = S_0 - \frac{2}{4!} \int d^{11} x \,\Omega \,\lambda_{abcd}  f^{(-) abcd} \label{Snew0} \,,\\
&S^\prime_{-3} = S_{-3} + \inv{4!}\int d^{11} x \,\Omega\,  \lambda_{abcd} \lambda^{abcd} \,.
\end{align}
\end{subequations}
Eliminating the auxiliary field via its equation of motion leads back to the original action. We see that after taking the limit that $c \to \infty$ the auxiliary field becomes a Lagrange multiplier imposing the constraint
\begin{equation}\label{constraintf}
 f^{(-)}_{ abcd}=0\,.
\end{equation}

The final non-relativistic action can now be obtained by determining the finite part $S_0^\prime$ and has been given in \cite{Blair:2021waq,Bergshoeff:2024nin} which we will not repeat here.

The limit described above is a critical limit in the sense that the cancellation of divergent $(T_{ab}^A)^2$ terms is due to a fine-tuning between the Einstein-Hilbert term and the kinetic term of the 3-form gauge potential. The same fine-tuning gives rise to an emergent local dilatation symmetry under which the Vierbeine transform as
\begin{equation}\label{localDM2}
\delta \tau_\mu^A = \lambda_D \tau_\mu^A\,,\hskip 2truecm \delta e_\mu^a = -\tfrac{1}{2}\lambda_D e_\mu^a\,,
\end{equation}
thereby generalizing a global scale symmetry triggered by the redefinitions \eqref{eq: M2-brane limit field redefinitions} to a local one. As explained in section 6 of \cite{Bergshoeff:2023rkk}, using a first-order formalism of general relativity the non-relativistic Einstein-Hilbert term, which is proportional to the curvature of spatial rotations, is invariant under these local dilatations by itself. This is due to the fact that the pre-factor $\Omega e^\mu_a e^\nu_b$ multiplying the spatial curvature $R_{\mu\nu}{}^{ab}(J)$ is invariant under the local dilatations \eqref{localDM2} for $D=11$ and membrane foliations. Using a second-order formulation instead, leads to the same non-relativistic expression plus a term
proportional to $(T_{aA}^A)^2$ that breaks local dilatations.\footnote{See eq.~(6.48) of  \cite{Bergshoeff:2023rkk}.} Remarkably, this dilatation-violating term is precisely cancelled by a similar contribution from the 3-form kinetic term thereby restoring the invariance under local dilatations of the final action.

The nice thing of the above argument is that it also applies to taking an M5-brane limit. Using a 5-brane foliation and requiring invariance of the same pre-factor  $\Omega e^\mu_a e^\nu_b$ now leads to the following  local dilatations:
\begin{equation}\label{localDM5}
\delta \tau_\mu^A = \lambda_D \tau_\mu^A\,,\hskip 2truecm \delta e_\mu^a = -2\lambda_D e_\mu^a\,.
\end{equation}
There is one difference however with the  membrane limit as we will see in the next section. The cancellation of the dilatation-violating term  $(T_{aA}^A)^2$  in the 5-brane limit does not follow from a straight contribution of the 3-form kinetic term but, instead, it follows as a by-product of applying a particular Hubbard-Stratonovich transformation. Thus the local invariance arises from a subtle cancellation between the metric and form degrees of freedom.

Before we close this section, we would like to point out that the self-duality constraint \eqref{constraintf} that we found in taking the M2-brane limit also follows from the duality relation \eqref{eq: relativistic duality relation} defining the dual 6-form, see also \cite{Blair:2021waq}.  As we are now using the dual 6-form field, we must supplement the redefinitions \eqref{eq: M2-brane limit field redefinitions} with the expansion\,\footnote{The $c^3$ term in this expansion is triggered by  the Chern-Simons term in the supergravity action. Without this Chern-Simons term, this $c^3$-term would be absent and the M2-brane limit of the duality relation will not lead to the self-duality constraint $f_{abcd}^{(-)}=0$ but instead would give $f_{abcd}=0$.}
\begin{equation}
    C_6 = \frac{c^3}{12}\epsilon_{ABC} \tau^A \wedge \tau^B \wedge \tau^C \wedge c_3 + c_6\label{eq: C6M2} \ ,
\end{equation}
that leads to the following covariant expansion of the curvatures:
\begin{subequations}
\begin{align}
    F_4 &= \frac{c^3}{2} \epsilon_{ABC} T^A \wedge \tau^B \wedge \tau^C + f_4 \ , \\
    G_7 &= - \frac{c^3}{3!} \epsilon_{ABC} \tau^A \wedge \tau^B \wedge \tau^C \wedge f_4  + g_7 \ ,
\end{align}
\end{subequations}
where $f_4 = dc_3$ and $g_7 = dc_6 - \tfrac{1}{2} c_3\wedge f_4$.

Since we have an M2-like foliation of our spacetime, we can contract with up to three longitudinal vectors and will therefore find four different projections. We immediately see that the divergent part of $G_7$ is only relevant when we contract with all three longitudinal vectors and can be ignored in the other cases, whereas the divergent part of $F_4$ can contribute to the equation when contracting with zero or one longitudinal vectors. For these two projections we explicitly find
\begin{subequations}
\begin{align}
    g_{a_1 ... a_7} &= - c^{-3} \epsilon_{a_1 ... a_7 b} T^B_{bB} + \frac{c^{-6}}{3!} \epsilon_{a_1 ... a_7 b} \epsilon_{B_1 B_2 B_3}  f^{b B_1 B_2 B_3} \ , \\
    g_{a_1 ... a_6 A} &= \inv{2} \epsilon_{a_1 ... a_6 b_1 b_2} \eta_{AB} T^B_{b_1 b_2} + \frac{c^{-3}}{4} \epsilon_{a_1 ... a_6 b_1 b_2} \epsilon_{A B_1 B_2} f^{b_1 b_2 B_1 B_2} \ ,
\end{align}
\end{subequations}
and so the divergent term of $F_4$ is still subleading in the totally transverse equation. The projection with two longitudinal vectors, in which neither divergent term contributes, takes the form
\begin{equation}
    g_{a_1 ... a_5 A_1 A_2 } = \inv{3!} \epsilon_{a_1 ... a_8}\epsilon_{A_1 A_2 A_3} f^{a_6 a_7 a_8 A_3} \ .
\end{equation}
Finally, the projection with all three longitudinal vectors is
\begin{equation}
    f_{a_1 ... a_4} + \inv{4!} \epsilon_{a_1 ... a_4 b_1 ... b_4} f_{b_1 ... b_4} = - \frac{c^{-3}}{3!} \epsilon^{B_1 B_2 B_3} g_{a_1 ... a_4 B_1 B_2 B_3} \ .
\end{equation}
Taking the $c\to\infty$ limit and collecting everything together, we find the equations
\begin{subequations}
\begin{align}
    g_{a_1 ... a_7} &= 0 \ , \\
    g_{a_1 ... a_6 A} &= \inv{2} \epsilon_{a_1 ... a_8} \eta_{AB} T^B_{a_7 a_8} \ , \\
    g_{a_1 ... a_5 A_1 A_2 } &= \inv{3!} \epsilon_{a_1 ... a_8}\epsilon_{A_1 A_2 A_3} f^{a_6 a_7 a_8 A_3} \ , \\
    f_{a_1 ... a_4} &= - \inv{4!} \epsilon_{a_1 ... a_8} f_{a_5 ... a_8} \ .
\end{align}
\end{subequations}
Three of these equations define the relevant components of $g_7$, but the final one gives a constraint on $f_4$; this is exactly the self-duality constraint that one finds from taking the M2-brane limit of the action.

\section{The M5-Brane Limit of the Action} \label{sect: M5-brane limit of action}

\subsection{Expansion of the Action}

Let us now turn our attention to the M5-brane limit of the theory, defined by making the field redefinitions (now $A=0,1,..,5$ and $a=6,7,8,9,10$)
\begin{subequations} \label{eq: M5-brane limit field redefinitions}
\begin{align}
    E^{\hat{A}}_{\mu} &= \brac{c \tau^A_{\mu}, c^{-2} e^a_{\mu}} \equiv (\hat{\tau}^A_{\mu}, \hat{e}^a_{\mu}) \ , \\
    C_3 &= c_3 \ , \\
    C_6 &= \frac{c^6}{6!} \epsilon_{A_1 ... A_6} \tau^{A_1} \wedge ... \wedge \tau^{A_6} + c_6 \ ,\label{eq: C6is}
\end{align}
\end{subequations}
before taking $c\to\infty$. With these the field strengths become
\begin{subequations}
\begin{align}
    F_4 &= f_4 \ , \\
    G_7 &= \frac{c^6}{5!} \epsilon_{A_1 ... A_6} T^{A_1} \wedge \tau^{A_2} \wedge ... \wedge \tau^{A_6} + g_7 \ ,
\end{align}
\end{subequations}
where we have defined $f_4 = dc_3$, $g_7 = dc_6 - \inv{2} c_3 \wedge f_4$, and $T^A = d\tau^A$. The limit of the action in this case is conceptually different to that of the M2-brane limit, as the only form-field in the action is $C_3$, which is unchanged by \eqref{eq: M5-brane limit field redefinitions}. However, we shall see that we still find a well-defined theory.

Let us consider the divergent terms in the action. The Ricci scalar has the expansion \cite{Bergshoeff:2024ipq}
\begin{subequations} \label{eq: ricci scalar expansion}
\begin{align}
    R &= c^4 \brac{c^6 R^{(6)} + R^{(0)} + O(c^{-6}) } \ , \\
    R^{(6)} &= - \inv{4} \eta_{AB} T_{ab}^A T_{ab}^B \ ,
\end{align}
\end{subequations}
where we will group all finite terms together as $R^{(0)}$ for now. The Chern-Simons term is manifestly finite throughout, so the only other concern is the kinetic term for $F_4$, which has the expansion
\begin{equation}
    S_{F,k} = - \inv{96 \kappa_{11}^2} \bigintssss d^{11} x \, \Omega \brac{
    c^{12} f_{abcd} f^{abcd} + 4 c^6 f_{abc A} f^{abc A} + 6 f_{ab AB} f^{ab AB} + O(c^{-6}) } \ ,
\end{equation}
where $\Omega$ is the non-relativistic measure given by the volume form
\begin{equation}
    \varepsilon_{\Omega} = \inv{5!6!} \epsilon_{ABCDEF} \epsilon_{abcde} \tau^A\wedge ... \wedge \tau^F \wedge e^a \wedge ... \wedge e^e \ .
\end{equation}
With this definition, the derivative of $\Omega$ is
\begin{equation} \label{eq: derivative of measure}
    \partial_{\mu} \ln \Omega = e^{\nu}_a \partial_{\mu} e_{\nu}^a + \tau^{\nu}_A \partial_{\mu} \tau_{\nu}^A \ .
\end{equation}
Our action is then
\begin{subequations}
\begin{align}
    S_{11d} &= c^{12} S_{12} + c^6 S_{6} + S_{0} + O(c^{-6}) \ , \\
    S_{12} &= - \inv{96 \kappa_{11}^2} \bigintssss d^{11}x \,\Omega \, f_{abcd} f^{abcd} \ , \\
    S_6 &= - \inv{4 \kappa_{11}^2} \bigintssss d^{11} x \, \Omega \brac{ \inv{2} \eta_{AB} T^A_{ab} T^B_{ab} + \inv{3 !} f_{abc A} f^{abc A} } \ , \\
    S_0 &= \inv{2 \kappa_{11}^2} \bigintssss d^{11}x\,\Omega \brac{
    R^{(0)} - \inv{8} f_{abAB} f^{abAB} } - \inv{12 \kappa_{11}^2} \bigintssss c_3 \wedge f_4 \wedge f_4 \ .
\end{align}
\end{subequations}
Our main interest is in the $O(c^6)$ terms, which can be rewritten as
\begin{align} \nonumber
    S_6 = - \inv{4\kappa_{11}^2} \bigintssss d^{11}x \, \Omega \bigg(
    &\inv{2} \eta_{AB} \brac{T_{ab}^A \mp \inv{3!} \epsilon_{abcde} f^{cde A} } \brac{T_{ab}^B \mp \inv{3!} \epsilon_{abcde} f^{cde B} } \\
    &\pm \inv{3!} \epsilon_{abcde} T^A_{ab} f_{cdeA} \bigg) \ .
\end{align}
We would like to use Hubbard-Stratonovich transformations to regulate the two divergent contributions to the action. However, the cross-term in the $O(c^6)$ action provides an impediment to this and must be dealt with. To do this, we use the totally transverse projection of the 3-form field's Bianchi identity,
\begin{equation}
    0 = \epsilon_{a_1 ... a_5} e^{\mu_1}_{a_1} ... e^{\mu_5}_{a_5} \partial_{[\mu_1} f_{\mu_2 ... \mu_5]} \ .
\end{equation}
Using the orthonormality of our basis vectors and vielbeins, and the completeness relation $\delta^{\mu}_{\nu} = \tau_A^{\mu} \tau^A_{\nu} + e^{\mu}_a e^a_{\nu}$, we can bring the vector fields inside the derivative to find
\begin{align} \nonumber
    0 = \ &\epsilon_{a_1 ... a_5} \bigg[ \partial_{\mu} \brac{e^{\mu}_{a_1} f_{a_2 ... a_5} }  + e^{\mu}_{a_1} \partial_{\mu} \ln\Omega \, f_{a_2 ... a_5} - T^A_{a_1 A} f_{a_2 ... a_5} \\ \label{eq: transverse bianchi}
    &- 2 T^A_{a_1 a_2} f_{a_3 ... a_5 A} - \inv{2} E^b_{[a_1 a_2} f_{a_3 ... a_5 b]} \bigg] \ ,
\end{align}
where we have used \eqref{eq: derivative of measure} and the notation
\begin{equation}
    E^a_{\mn} = 2 \partial_{[\mu} e^a_{\nu]} \ .
\end{equation}
As the final term is antisymmetrised over six transverse indices it must vanish. It will be convenient to use the non-relativistic connection
\begin{equation} \label{eq: NR connection}
    \Gamma^{\rho}_{\mn} = \inv{2} H^{\rho \sigma} \brac{\partial_{\mu} H_{\nu\sigma} + \partial_{\nu} H_{\mu\sigma} - \partial_{\sigma} H_{\mn}} + \tau^{\rho}_A \partial_{\mu} \tau_{\nu}^A \ ,
\end{equation}
and so the divergence of a vector field $X$ is given explicitly by
\begin{equation} \label{eq: NR divergence}
    \nabla_{\mu} X^{\mu} = \partial_{\mu} X^{\mu} + X^{\mu} \partial_{\mu} \ln\Omega - T^A_{\mu A} X^{\mu} \ .
\end{equation}
Comparing this to the first three terms in \eqref{eq: transverse bianchi}, we find the identity
\begin{equation} \label{eq: cross term identity}
    \epsilon_{abcde} T^A_{ab} f_{cdeA} = \inv{2} \nabla_{\mu} \brac{\epsilon_{abcde} e^{\mu}_a f_{bcde}} \ .
\end{equation}
From \eqref{eq: NR divergence} we see that the divergence theorem for our torsional connection is
\begin{equation}
    \int d^{11}x \, \Omega \, \nabla_{\mu} X^{\mu} = - \int d^{11}x \, \Omega \, T_{\mu A}^A X^{\mu}\ ,
\end{equation}
when boundary terms can be ignored, and so the cross-term can be brought into the form
\begin{equation}
    \int d^{11}x \, \Omega \, \epsilon_{abcde} T^A_{ab} f_{cdeA} = - \inv{2} \int d^{11}x \, \Omega \, \Tilde{T}_{abcd} f_{abcd}  \ ,
\end{equation}
where we have defined
\begin{equation} \label{eq: L 4-index definition}
    \Tilde{T}_{abcd} = \epsilon_{abcde} T^A_{eA} \ .
\end{equation}
The $O(c^6)$ action can then be rewritten as
\begin{align} \nonumber
    S_6 = - \inv{4\kappa_{11}^2} \bigintssss d^{11}x \, \Omega \bigg( &
    \inv{2} \eta_{AB} \brac{T_{ab}^A \mp \inv{3!} \epsilon_{abcde} f^{cde A} } \brac{T_{ab}^B \mp \inv{3!} \epsilon_{abcde} f^{cde B} } \\
    &\mp \frac{1}{12} \Tilde{T}_{abcd} f_{abcd} \bigg) \ .
\end{align}
The final term can absorbed into $S_{12}$ and $S_0$, yielding
\begin{subequations}
\begin{align}
    S_{12}' &= - \inv{96 \kappa_{11}^2} \bigintssss d^{11}x \,\Omega \, \brac{f_{abcd} \mp c^{-6}  \Tilde{T}_{abcd}} \brac{f^{abcd} \mp  c^{-6} \Tilde{T}^{abcd} }\label{eq: S12prime} \ , \\
    S_0' &= \inv{2 \kappa_{11}^2} \bigintssss d^{11}x\,\Omega \brac{
    R^{(0)} - \inv{8} f_{abAB} f^{abAB} + \inv{48} \Tilde{T}_{abcd} \Tilde{T}^{abcd} } - \inv{12 \kappa_{11}^2} \bigintssss c_3 \wedge f_4 \wedge f_4 \ .
\end{align}    
\end{subequations}
The divergent terms are now in the form required for the Hubbard-Stratonovich transformations: performing these and taking the $c\to\infty$ limit yields
\begin{align} \nonumber
    S_{M5} = \inv{2\kappa_{11}^2} \bigintssss d^{11}x \, \Omega &\bigg(
    R^{(0)} + \lambda^{abcd} f_{abcd} + \Lambda^{ab}_A \brac{T^A_{ab} \mp \inv{3!} \epsilon_{abcde} f^{cdeA}} - \inv{8} f_{abAB} f^{abAB} \\ 
    &+ \inv{2} T_{a A}^A T_{a B}^B \bigg) - \inv{12 \kappa_{11}^2} \bigintssss c_3 \wedge f_4 \wedge f_4 \ .
\end{align}
Finally, using the explicit form of $R^{(0)}$ (see appendix \ref{sect: curvature expansion} for details) the action becomes
\begin{align} \nonumber
    S_{M5} = \inv{2\kappa_{11}^2} \bigintssss d^{11}x \, \Omega &\bigg(
    H^{\mn} \cal{R}_{\mn} -2 H^{\mn} \tau^{\rho}_A \nabla_{\mu} T_{\nu\rho}^A - \inv{2} \big( T^A_{aB} T^B_{aA} + \eta_{AB} \eta^{CD} T_{aC}^A T_{aD}^B \\ \nonumber
    &+ T^{A}_{aA}T^B_{aB} \big) +  \lambda^{abcd} f_{abcd} + \Lambda^{ab}_A \brac{T^A_{ab} \mp \inv{3!} \epsilon_{abcde} f^{cdeA}}  \\ \label{eq: M5 limit of action}
    &- \inv{8} f_{abAB} f^{abAB} \bigg) - \inv{12 \kappa_{11}^2} \bigintssss c_3 \wedge f_4 \wedge f_4 \ .
\end{align}
The Lagrange multiplier fields impose the two constraints
\begin{subequations}
\begin{align}
    f_{abcd} &= 0 \ , \\
    T^A_{ab} &= \pm \inv{3!} \epsilon_{abcde} f^{cdeA} \ .\label{eq: 2ndconstraint}
\end{align}
\end{subequations}
We note that there is a sign ambiguity in the action formulation stemming from the fact that the action is blind to $C_6$, and therefore cannot differentiate between an M5-brane limit and an anti-M5-brane limit ({\it i.e.} between a plus and a minus in front of the divergent term in $C_6$). Throughout this work we will take the upper sign and work with the M5-brane limit.

\subsection{Symmetries of the Action}

Let us look at the symmetries of the action \eqref{eq: M5 limit of action}. As with the relativistic theory, it is obviously invariant under spacetime diffeomorphisms. As we are working with objects built from vielbeins, we also have the remnant of the local $SO(1,10)$ symmetry that survives the M5-brane limit. The symmetry-breaking pattern $SO(1,10) \to SO(1,5) \times SO(5)$ imposed by the M5-brane limit means we expect longitudinal local Lorentz transformations and transverse Euclidean rotations to still be symmetries of the theory. The infinitesimal form of these transformations is
\begin{subequations}
\begin{align}
    \delta \tau^A_{\mu} &= \tensor{\omega}{^A_B} \tau^B_{\mu} \ , \\
    \delta e^a_{\mu} &= \tensor{r}{^a_b} e_{\mu}^b \ , \\
    \delta \lambda^{abcd} &= 4\tensor{r}{^{[a}_e} \lambda^{|e|bcd]} \ , \\
    \delta \Lambda^{ab}_A &= 2 \tensor{r}{^{[a}_c} \Lambda^{|c|b]}_A - \tensor{\omega}{^B_A} \Lambda^{ab}_B \ , \\
    \delta c_3 &= 0 \ ,
\end{align}
\end{subequations}
where $\tensor{\omega}{^A_B}(x)$ and $\tensor{r}{^a_b}(x)$ are arbitrary well-behaved functions satisfying
\begin{subequations}
\begin{align}
    \tensor{\omega}{_{(AB)}} &= 0 \ , \\
    \tensor{r}{_{(ab)}} &= 0 \ .
\end{align}
\end{subequations}
While the Euclidean rotations are obviously symmetries, it is less clear that this is true for the Lorentzian transformations due to the presence of $\tau^A$ derivatives in $\Gamma^{\mu}_{\nu\rho}$ and $T^A$. Using the explicit expressions
\begin{subequations}
\begin{align}
    \delta T^A_{\mn} &= \tensor{\omega}{^A_B} T^B_{\mn} + 2 \partial_{[\mu} \tensor{\omega}{^A_B} \tau^B_{\nu]}  \ , \\
    \delta \Gamma^{\rho}_{\mn} &= \tau^{\rho}_A \partial_{\mu} \tensor{\omega}{^A_B} \tau^B_{\nu} \ ,
\end{align}
\end{subequations}
in consort with \eqref{eq: Ricci tensor variation} we see that the first two terms in \eqref{eq: M5 limit of action} are invariant, but the third and fourth transform as
\begin{subequations}
\begin{align}
    \delta \brac{T^A_{aB} T^B_{aA} } &= 2 e^{\mu}_a \partial_{\mu} \tensor{\omega}{^A_B} T^B_{aA} \ , \\
    \delta \brac{\eta_{AB} \eta^{CD} T^A_{aC} T^B_{aD}} &= - 2 e^{\mu}_a \partial_{\mu} \tensor{\omega}{^A_B} T^B_{aA} \ .
\end{align}
\end{subequations}
Thankfully, while neither are individually invariant the sum of the two is, and we see that the action is invariant.

We also have the limit of the relativistic transformations mixing $\tau^A$ and $e^a$, which become the 5-brane Galilean boost transformations
\begin{subequations}
\begin{align}
    \delta \tau^A_{\mu} &= 0 \ , \\
    \delta \tau^{\mu}_A &= - v^a_A e_a^{\mu} \ , \\
    \delta e^a_{\mu} &=  v^a_A \tau_{\mu}^A \ , \\
    \delta e_a^{\mu} &= 0 \ ,
\end{align}
\end{subequations}
for a set of functions $v^a_A(x)$. Using the relativistic transformations we see that the boosts leaves the three-form field invariant. Using this, it is straightforward (but tedious) to show that the transformation of the action is
\begin{align} \nonumber
    \delta S_{M5} = \inv{2\kappa_{11}^2} \int d^{11} x \, \Omega \bigg[ &
    - \eta_{AB} T^A_{ab} Y^B_{ab} - \inv{3!} f_{abcd} v_a^A \epsilon_{bcdef} \Lambda^{ef}_A - \inv{2} v^B_c f_{abc A} f_{abAB} \\ \label{eq: boost transformed action 1}
    &+ \delta \lambda_{abcd} f_{abcd} + \delta \Lambda^{ab}_A \brac{T^A_{ab} - \inv{3!} \epsilon_{abcde} f^{cdeA} } \bigg] \ ,
\end{align}
where $Y^A_{ab}$ is given by
\begin{equation}
    Y^A_{ab} = e^{\mu}_{[a} \partial_{|\mu|} v^A_{b]} - v^A_{[a} T^B_{b] B} + v^B_{[a} T^A_{b]B} + \inv{2} v^A_{c} E^c_{ab} \ .
\end{equation}
By taking the second Lagrange multiplier field to have the transformation
\begin{equation}
    \delta \Lambda^{ab}_A = Y^{ab}_A - \inv{4} \epsilon_{abcde} v^B_c f_{deAB} \ ,
\end{equation}
we can cancel the first and third terms in \eqref{eq: boost transformed action 1} at the expense of the cross-terms
\begin{align} \nonumber
    \cal{C} &= - \inv{24} \epsilon_{abcde} \brac{4 Y^A_{ab} f_{cdeA} + 6 \epsilon_{abcde} T^A_{ab} v^B_c f_{deAB} } \\ \nonumber
    &\equiv \inv{24} \epsilon_{abcde} v_e^A \bigg[ 6 T^B_{ab} f_{cdAB} + 4 T^B_{aB} f_{bcdA} + 4 T^B_{Aa} f_{bcdB} \bigg] \\
    &\qquad - \inv{12} \epsilon_{abcde} E^f_{ab} f_{cdeA} v^A_f + \inv{6} \epsilon_{abcde} e^{\mu}_a f_{bcdA} \partial_{\mu} v^A_e  \ .
\end{align}
To proceed, we must make use of the projection
\begin{equation}
    0 = \tau^{\mu_1}_A e^{\mu_2}_a ... e^{\mu_5}_d \partial_{[\mu_1} f_{\mu_2 ... \mu_5]}\ ,
\end{equation}
of the Bianchi identity. By taking the basis vectors inside the derivative, we see that this is equivalent to the identity
\begin{align} \nonumber
    0 = \ & \nabla_{\mu} \brac{\tau^{\mu}_A f_{abcd}} + 4 \nabla_{\mu} \brac{e^{\mu}_{[a} f_{bcd]A}} + E^e_{eA} f_{abcd} + 4 E^e_{e[a} f_{bcd]A} - 4 T^B_{A[a} f_{bcd]B} \\
    &- 4 E^e_{A[a} f_{bcd]e} - 6 T^B_{[ab} f_{cd]AB} + 6 E^e_{[ab} f_{cd]eA} \ .
\end{align}
The cross terms can then be written as
\begin{align} \nonumber
    \cal{C} = \ &\inv{24} \epsilon_{abcde} \Bigg[
    \nabla_{\mu} \brac{v_e^A \tau^{\mu}_A f_{abcd}} - \tau^{\mu}_A f_{abcd} \partial_{\mu} v_e^A + 4 \nabla_{\mu} \brac{v^A_e e^{\mu}_{a} f_{bcdA} } \\ \nonumber 
    &+ 4 v^A_e T^B_{aB} f_{bcdA} + v_e^A E^f_{fA} f_{abcd} + 4 v_e^A E^f_{f a} f_{bcdA} - 4 v_e^A E^f_{A a} f_{bcdf} \\
    &+ 6 v^e_A E^f_{ab} f_{cdfA} \bigg] - \inv{12} \epsilon_{abcde} E^f_{ab} f_{cdeA} v^A_f \ .
\end{align}
The three terms involving totally transverse projections of $E^a$ vanish due to the identity
\begin{align} \nonumber
    0 &= 6\epsilon_{abcde} E^f_{[ab} f_{cde|A|} v^A_{f]} \\
    &= \epsilon_{abcde} \brac{
    E^f_{ab} f_{cdeA} v^A_f - 3 E^f_{ab} f_{cdfA} v^A_e - 2 E^f_{fa} f_{bcdA} v^A_e } \ ,
\end{align}
and we can simplify the other two $E^a$ terms using
\begin{align} \nonumber
    0 &= 6 \epsilon_{abcde} E^f_{A[a} f_{bcde} v^A_{f]} \\
    &= \epsilon_{abcde} \brac{E^f_{Aa} f_{bcde} v^A_f - 4 E^f_{Aa} f_{bcdf} v_e^A - E^f_{Af} f_{abcd} v_e^A} \ .
\end{align}
Using the divergence theorem \eqref{eq: NR divergence} we can integrate the remaining cross terms to find
\begin{equation}
    \int d^{11} x \, \Omega \, \cal{C} = - \frac{\epsilon_{abcde}}{24} \int d^{11}x \, \Omega \, f_{abcd} \brac{ \tau_A^{\mu} \partial_{\mu} v^A_e + T^B_{AB} v^A_e + E^f_{Ae} v^A_f } \ ,
\end{equation}
and so we see that the transformation of the action is
\begin{align} \nonumber
    \delta S_{M5} = \inv{2\kappa_{11}^2} \int d^{11} x \, \Omega \, f_{abcd} \bigg[&
    \delta \lambda_{abcd} - \inv{6} v^A_a \epsilon_{bcdef} \Lambda^{ef}_A -  \frac{\epsilon_{abcde}}{24} \Big( \tau_A^{\mu} \partial_{\mu} v^A_e \\
    &+ T^B_{AB} v^A_e + E^f_{Ae} v^A_f \Big) \bigg] \ .
\end{align}
Hence, by taking the fields to have the transformations
\begin{subequations} \label{eq: boost transformations}
\begin{align}
    \delta \tau^A_{\mu} &= 0 \ , \\
    \delta \tau^{\mu}_A &= - v^a_A e_a^{\mu} \ , \\
    \delta e^a_{\mu} &=  v^a_A \tau_{\mu}^A \ , \\
    \delta e_a^{\mu} &= 0 \ , \\
    \delta c_3 &= 0 \ , \\
    \delta \Lambda_{ab}^A &= e^{\mu}_{[a} \partial_{|\mu|} v^A_{b]} - v^A_{[a} T^B_{b] B} + v^B_{[a} T^A_{b]B} + \inv{2} v^A_{c} E^c_{ab} - \inv{4} \epsilon_{abcde} v_B^c f^{deAB} \ , \\
    \delta \lambda_{abcd} &= \inv{6} v^A_a \epsilon_{bcdef} \Lambda^{ef}_A +  \frac{\epsilon_{abcde}}{24} \Big( \tau_A^{\mu} \partial_{\mu} v^A_e + T^B_{AB} v^A_e + E^f_{Ae} v^A_f \Big) \ ,
\end{align}
\end{subequations}
we see that the action is invariant under five-brane Galilean boosts.

Finally, \eqref{eq: M5 limit of action} is also invariant under the local dilatation transformation
\begin{subequations} \label{eq: local dilatation symmetry}
\begin{align}
    \Tilde{\tau}_{\mu}^A &= e^{\sigma} \tau_{\mu}^A \ , \\
    \Tilde{e}_{\mu}^a &= e^{-2\sigma} e_{\mu}^a \ , \\
    \Tilde{c}_{\mn\rho} &= c_{\mn\rho} \ , \\
    \Tilde{\lambda}^{abcd} &= e^{-4\sigma} \lambda^{abcd} \ , \\
    \Tilde{\Lambda}^{ab}_A &= e^{-\sigma} \Lambda_A^{ab} \ ,
\end{align}
\end{subequations}
with $\sigma(x)$ an arbitrary smooth function that parametrises the action of the scale transformation. In what follows we view this as a gauge symmetry. Imposing both constraints for the system, this implies the relation
\begin{equation} \label{eq: eom trace relation}
    2 e^{\mu}_a \bbm{E}^{a}_{\mu} + \tau_{\mu}^A \bbm{E}_{A}^{\mu} = 0
\end{equation}
between the traces of the equations of motion for $e^{\mu}_a$ and $\tau_{\mu}^A$. We therefore lose an equation of motion in the action formulation when performing the M5-brane limit. This will turn out to be the Poisson equation, and in the next section we shall see how it appears from a direct expansion of the relativistic equations of motion.

\section{The M5-Brane Limit of the Equations of Motion} \label{sect: equations of motion limit}

\subsection{The Duality Relation} \label{sect: duality relation limit}

In order to find the additional equation of motion missing from the action formulation we must perform a consistent limit of the relativistic equations of motion. We first examine the duality relation \eqref{eq: relativistic duality relation}, which in components is
\begin{equation} \label{eq: rel duality relation in comp}
    G_{\mu_1 ... \mu_7} = \inv{4!} \varepsilon_{\mu_1 ... \mu_7 \nu_1 ... \nu_4} g^{\nu_1 \rho_1} ... g^{\nu_4 \rho_4} F_{\rho_1 ... \rho_4} \ .
\end{equation}
In the M2-brane limit this was where the action's constraints were encoded in the equations of motion, and we shall see that the same is true for the M5-brane limit. The limit is most easily taken by considering all possible contractions of \eqref{eq: rel duality relation in comp} with the orthonormal basis $\{\hat{\tau}^{\mu}_A,\hat{e}^{\mu}_a\}$ of the relativistic metric, as with this normalisation the only non-vanishing contraction of the volume form is
\begin{equation}
    \varepsilon_{\mu_1 ... \mu_6 \nu_1 ... \nu_5} \hat{\tau}^{\mu_1}_{A_1} ... \hat{\tau}^{\mu_6}_{A_6} \hat{e}^{\nu_1}_{a_1} ... \hat{e}^{\nu_5}_{a_5} = \epsilon_{A_1 ... A_6} \epsilon_{a_1 ... a_5} \ ,
\end{equation}
before expanding in $c$ and isolating the leading-order term. 

As the duality relation is antisymmetrised, the only possible contractions are those with two to six longitudinal vectors. The divergent term in the expansion of $G_7$ is only non-zero when contracted with five or six longitudinal vectors, and so it is easy to see that the first three projections give
\begin{subequations}
\begin{align}
    g_{a_1...a_5 A_1 A_2} &= \frac{c^{-12}}{4!} \epsilon_{a_1 ... a_5} \epsilon_{A_1 A_2 B_1 ... B_4} f^{B_1 ... B_4} \ , \\
    g_{a_1 ... a_4 A_1 A_2 A_3} &= - \frac{c^{-6}}{3!} \epsilon_{a_1 ... a_4 b} \epsilon_{A_1 A_2 A_3 B_1 B_2 B_3} f^{b B_1 B_2 B_3} \ , \\
    g_{a_1 a_2 a_3 A_1 ... A_4} &= \inv{4} \epsilon_{a_1 a_2 a_3 b_1 b_2} \epsilon_{A_1 ... A_4 B_1 B_2} f^{b_1 b_2 B_1 B_2} \ .
\end{align}
\end{subequations}
For the final two, we also have the contribution from the divergent term. When contracting with five longitudinal vectors we find
\begin{align} \nonumber
    T^A_{a_1 a_2} &= \inv{3!} \epsilon_{a_1 a_2 b_1 b_2 b_3 } f^{b_1 b_2 b_3 A} + \frac{c^{-6}}{5!} \epsilon^{AB_1 ... B_5} g_{a_1 a_2 B_1 ... B_5} \\ \label{eq: duality constraint 1}
    &\equiv \inv{3!} \epsilon_{a_1 a_2 b_1 b_2 b_3 } f^{b_1 b_2 b_3 A} + c^{-6} \Tilde{g}^A_{a_1 a_2} \ ,
\end{align}
and when contracting with all six we have
\begin{align} \nonumber
    f_{abcd} &= \epsilon_{abcde} \brac{
    c^{-6} T^A_{eA} - \frac{c^{-12}}{6!} \epsilon^{B_1 ... B_6} g_{e B_1 ... B_6} } \\ \label{eq: duality constraint 2}
    &\equiv c^{-6} \Tilde{T}_{abcd} + c^{-12} \Tilde{g}_{abcd} \ .
\end{align}
While the subleading terms will be essential later on, they play no role in the equations of motion derived from the duality relation. Putting this all together, the equations we find in the $c\to\infty$ limit are
\begin{subequations} \label{eq: M5 limit of duality relation}
\begin{align}
    g_{a_1 ... a_5 A_1 A_2} &= 0 \ , \\
    g_{a_1 ... a_4 A_1 A_2 A_3} &= 0 \ , \\ \label{eq: NR Duality equation}
    g_{a_1 a_2 a_3 A_1 ... A_4} &= \inv{4} \epsilon_{a_1 ... a_5} \epsilon_{A_1 ... A_6} f^{a_4 a_5 A_5 A_6} \ , \\
    T_{a_1 a_2}^A &= \inv{3!} \epsilon_{a_1 ... a_5} f^{a_3 a_4 a_5 A} \ , \\
    f_{a_1 ... a_4} &= 0 \ .
\end{align}
\end{subequations}
While the first three of these can be considered definitions of the relevant components of $g_7$, the latter two are exactly the constraints imposed by the Lagrange multipliers in the action formulation of the non-relativistic theory.

\subsection{From the Einstein Equation to the Poisson Equation} \label{sect: einstein equation limit}

We shall now apply the M5-brane limit to the Einstein equation. Using the expansion of the Ricci curvature performed in appendix \ref{sect: curvature expansion}, its projections along the basis vectors is
\begin{subequations}
\begin{align}
    e^{\mu}_a e^{\nu}_b R_{\mn} &= - \frac{c^6}{2} \eta_{AB} T^A_{ac} T^B_{bc} + e^{\mu}_a e^{\nu}_b R_{\mn}^{(0)} + O(c^{-6}) \ , \\
    e^{\mu}_a \tau^{\nu}_A R_{\mn} &= \frac{c^6}{2} \brac{
    \eta_{AB} e^{\mu}_a H^{\rho\sigma} \nabla_{\rho} T_{\mu\sigma}^B - 2\eta_{B(C}  T^C_{A)b} T^B_{ab} } + e^{\mu}_a \tau^{\nu}_A R_{\mn}^{(0)} + O(c^{-6}) \ , \\ \nonumber
    \tau^{\mu}_A \tau^{\nu}_B R_{\mn}  &=  \, \frac{c^{12}}{4} \eta_{AC} \eta_{BD} T^C_{ab} T^D_{ab} + c^6 \bigg[
    \tau^{\mu}_{(A} \eta_{B)C} H^{\nu\rho} \nabla_{\nu } T_{\mu\rho}^C - T^D_{aD} T^C_{a(A} \eta_{B)C} \\
    &\qquad + \inv{2} \brac{
    \eta_{AC} \eta_{BD} \eta^{EF} T^C_{aE} T^D_{aF} - \eta_{CD} T^C_{aA} T^D_{aB} } \bigg] + O(c^0) \ ,
\end{align}
\end{subequations}
where we have denoted the $c$-independent terms in $R_{\mn}$ by $R^{(0)}_{\mn}$. From this, we see that $R^{(0)}$ as defined in equation \eqref{eq: ricci scalar expansion} is
\begin{equation} \label{eq: finite bit of R}
    R^{(0)} = H^{\mn} R^{(0)}_{\mn} + \tau^{\mu}_A H^{\nu\rho} \nabla_{\nu} T^A_{\mu\rho} - T^A_{aA} T^B_{aB} \ ,
\end{equation}
and so the relativistic Einstein tensor has the expansion
\begin{subequations}
\begin{align} \nonumber
    e^{\mu}_a e^{\nu}_b G_{\mn} &= - \frac{c^6}{2} \eta_{AB} \brac{T^A_{ac} T^B_{bc} - \inv{4} \delta_{ab} T^A_{cd} T^B_{cd}  } + e^{\mu}_a e^{\nu}_b R_{\mn}^{(0)}- \inv{2} \delta_{ab} H^{\mn} R^{(0)}_{\mn}  \\
    &\qquad - \inv{2} \delta_{ab}\tau^{\mu}_A H^{\nu\rho} \nabla_{\nu} T^A_{\mu\rho} +\inv{2} \delta_{ab} T^A_{cA} T^B_{cB} +  O(c^{-6}) \ , \\ 
    e^{\mu}_a \tau^{\nu}_A G_{\mn} &= \frac{c^6}{2} \brac{
    \eta_{AB} e^{\mu}_a H^{\rho\sigma} \nabla_{\rho} T_{\mu\sigma}^B - 2\eta_{B(C}  T^C_{A)b} T^B_{ab} }  + e^{\mu}_a \tau^{\nu}_A R_{\mn}^{(0)} + O(c^{-6}) \ , \\ \nonumber
    \tau^{\mu}_A \tau^{\nu}_B G_{\mn}  &= \frac{c^{12}}{4} \brac{ \eta_{AC} \eta_{BD} T^C_{ab} T^D_{ab} + \inv{2} \eta_{AB} \eta_{CD} T^C_{ab} T^D_{ab} }  + c^6 \bigg[
    \tau^{\mu}_{(A} \eta_{B)C} H^{\nu\rho} \nabla_{\nu } T_{\mu\rho}^C \\ \nonumber
    &\qquad - \inv{2} \eta_{AB} \tau^{\mu}_C H^{\nu\rho} \nabla_{\nu} T^C_{\mu\rho} + \inv{2} \brac{
    \eta_{AC} \eta_{BD} \eta^{EF} T^C_{aE} T^D_{aF} - \eta_{CD} T^C_{aA} T^D_{aB} } \\
    &\qquad - T^D_{aD} T^C_{a(A} \eta_{B)C} + \inv{2} \eta_{AB} T^C_{aC} T^D_{aD} - \inv{2} \eta_{AB} H^{\mn}R^{(0)}_{\mn} \bigg] + O(c^0) \ .
\end{align}
\end{subequations}
Doing the same for the stress tensor gives
\begin{subequations}
\begin{align} \nonumber
    e^{\mu}_a e^{\nu}_b \Theta_{\mn} &= \inv{12} \bigg[
    c^{12} \brac{f_{acde} f_{bcde} - \inv{8} \delta_{ab} f_{cdef} f_{cdef} } + c^6 \Big(
    3 \eta^{AB} f_{acdA} f_{bcdB} \\ \nonumber
    & \qquad- \inv{2} \delta_{ab} \eta^{AB} f_{cdeA} f_{cdeB} \Big)  + 3 \eta^{AB} \eta^{CD} f_{acAC} f_{bcBD} \\
    & \qquad- \frac{3}{4} \delta_{ab} \eta^{AB} \eta^{CD} f_{cdAC} f_{cdBD} \bigg] + O(c^{-6}) \ , \\ 
    e^{\mu}_a \tau^{\nu}_A \Theta_{\mn} &= \inv{12} \bigg[
    c^{12} f_{abcd} f_{Abcd} + 3 c^6 \eta^{BC} f_{abcB} f_{AbcC}  + 3 \eta^{BC} \eta^{DE} f_{abBD} f_{AbCE} 
    \bigg] + O(c^{-6}) \ , \\ \nonumber
    \tau^{\mu}_A \tau^{\nu}_B \Theta_{\mn} &= \inv{12} \bigg[
    - \inv{8} c^{18} \eta_{AB} f_{abcd} f_{abcd} + c^{12} \brac{
    f_{Aabc} f_{Babc} - \inv{2} \eta_{AB} \eta^{CD} f_{Cabc} f_{Dabc}
    } \\
    &\qquad+ c^6 \brac{3 \eta^{CD} f_{ACab} f_{BDab} - \frac{3}{4} \eta_{AB} \eta^{CD} \eta^{EF} f_{CEab} f_{DFab} }
    \bigg] + O(c^0) \ .
\end{align}
\end{subequations}

We must now use the equations \eqref{eq: duality constraint 1} and \eqref{eq: duality constraint 2} derived from the duality relations. If we first consider the totally transverse projection, we see that the $O(c^6)$ terms in the Einstein tensor can be rewritten as
\begin{align} \nonumber
    - \frac{c^6}{2}\eta_{AB} \brac{T^A_{ac} T^B_{bc} - \inv{4} \delta_{ab} T^A_{cd} T^B_{cd}} =& \ \frac{c^6}{4} \eta^{AB} \brac{f_{acdA} f_{bcdB} - \inv{6} \delta_{ab} f_{cdeA} f_{cdeB} } \\
    &\ -  \eta_{AB} \brac{T^A_{(a|c|} \Tilde{g}^B_{b)c} - \inv{4} \delta_{ab} T^A_{cd} \Tilde{g}^B_{cd}} + O(c^{-6}) \ ,
\end{align}
with the $O(c^6)$ terms now identical to those in the expansion of the stress tensor. Similarly, the $O(c^{12})$ terms in the stress tensor become
\begin{align} \nonumber
    \frac{c^{12}}{12} \brac{f_{acde} f_{bcde} - \inv{8} \delta_{ab} f_{cdef} f_{cdef} } &= \inv{12} \brac{\Tilde{T}_{acde} \Tilde{T}_{bcde} - \inv{8} \delta_{ab} \Tilde{T}_{cdef} \Tilde{T}_{cdef}} + O(c^{-6}) \\
    &= - \inv{2} \brac{ T^A_{aA} T^B_{bB} - \inv{2} \delta_{ab} T^A_{cA} T^B_{cB} } + O(c^{-6}) \ .
\end{align}
The transverse projection of the Einstein equation is therefore
\begin{align} \nonumber
    0 = \ & e^{\mu}_a e^{\nu}_b R^{(0)}_{\mn} - \eta_{AB} T^A_{(a|c|} \Tilde{g}^B_{b)c} + \inv{2} T^A_{aA} T^B_{bB} - 3 \eta^{AB} \eta^{CD} f_{acAC} f_{bcBD} \\ \nonumber
    &- \inv{4}\delta_{ab} \bigg( 2 H^{\mn} R^{(0)}_{\mn} -
    \eta_{AB} T^A_{cd} \Tilde{g}^B_{cd} - T^A_{cA} T^B_{cB} + 2 \tau^{\mu}_A H^{\nu\rho} \nabla_{\nu} T^A_{\mu\rho} \\
    &+ 3 \eta^{AB} \eta^{CD} f_{cdAC} f_{cdBD} \bigg) + O(c^{-6}) \ .
\end{align}
An analogous computation for the mixed projection (where we must also divide through by a factor of $c^6$) gives
\begin{equation}
    0 = \eta_{AB} e^{\mu}_a H^{\rho\sigma} \nabla_{\rho} T^B_{\mu\sigma} + \inv{6} \epsilon_{abcde} T^B_{bA} f_{cdeB} - \inv{2} \eta^{BC} f_{abcB} f_{AbcC} + O(c^{-6}) \ .
\end{equation}

To find the totally longitudinal projection we must do slightly more work. Dividing both sides by a factor of $c^6$, the divergent terms arising in the Einstein tensor can be written as
\begin{align} \nonumber
    \frac{c^6}{4} \brac{\eta_{AC} \eta_{BD} T^C_{ab} T^D_{ab} + \inv{2} \eta_{AB} \eta_{CD} T^C_{ab} T^D_{ab} } = \ & \frac{c^6}{12} \brac{f_{abcA} f_{abcB} + \inv{4} \eta_{AB} \epsilon_{abcde} T^C_{ab} f_{cdeC} } \\
    &+ \inv{8} \eta_{AB} \eta_{CD} T^C_{ab} \Tilde{g}^D_{ab} \ ,
\end{align}
while the divergent terms in the stress tensor take the form
\begin{subequations}
\begin{align}
    - \frac{c^{12}}{96} \eta_{AB} f_{abcd} f_{abcd} &= - \inv{4} \eta_{AB} T^C_{aC} T^D_{aD} + O(c^{-6}) \ , \\ \nonumber
    \frac{c^6}{12} \brac{f_{abcA} f_{abcB} - \inv{2} \eta_{AB} \eta^{CD} f_{abcC} f_{abcD}} &= \frac{c^6}{12} \brac{f_{abcA} f_{abcB} - \inv{4} \eta_{AB} \epsilon_{abcde} T^C_{ab} f_{cdeC}} \\
    & \quad - \inv{4}\eta_{AB} \brac{T^C_{aC} T^D_{aD} - \inv{2} \eta_{CD} T^C_{ab} \Tilde{g}^D_{ab}} + O(c^{-6}) \ .
\end{align}
\end{subequations}
The divergent terms now do not fully cancel, leaving us with
\begin{equation}
    c^{-6} \tau^{\mu}_A \tau^{\nu}_B \brac{G_{\mn} - \Theta_{\mn} } = \frac{c^6}{24} \eta_{AB} \epsilon_{abcde} T^C_{ab} f_{cdeC} + O(c^0) \ .
\end{equation}
We recall that this combination of terms satisfies the identity \eqref{eq: cross term identity}, which when combined with equation \eqref{eq: duality constraint 2} shows us that
\begin{equation}
    \frac{c^6}{24} \eta_{AB} \epsilon_{abcde} T^C_{ab} f_{cdeC} = \eta_{AB} \nabla_{\mu} \brac{e^{\mu}_a T^C_{aC}} + c^{-6} \eta_{AB} \nabla_{\mu} g^{\mu} \ ,
\end{equation}
where we have defined the vector field
\begin{equation}
    g^{\mu} = - \inv{6!} e^{\mu}_a \epsilon^{ABCDEF} g_{aABCDEF} \ ,
\end{equation}
leaving us with only finite terms. After a brief calculation we find that the finite piece can be brought to the form
\begin{equation}
    \nabla_{\mu} \brac{e^{\mu}_a T^C_{aC}} = - \tau^{\mu}_C H^{\nu\rho} \nabla_{\nu} T^C_{\mu\rho} \ ,
\end{equation}
and so the totally longitudinal projection becomes
\begin{align} \nonumber
    0 = \ & \tau^{\mu}_{(A} \eta_{B)C} H^{\nu\rho} \nabla_{\nu} T^C_{\mu\rho} - T^D_{aD} T^C_{a(A} \eta_{B)C} + \inv{2} \brac{\eta_{AC} \eta_{BD} \eta^{EF} T^C_{aE} T^D_{aF} - \eta_{CD} T^C_{aA} T^D_{aB} }  \\ \nonumber
    &+ \inv{2} \eta_{(A|C|} \eta_{B)D} T^C_{ab} \Tilde{g}^D_{ab} -\inv{4} \eta^{CD} f_{ACab} f_{BDab} - \inv{2} \eta_{AB} \bigg(
    H^{\mn} R^{(0)}_{\mn} + 2 \tau^{\mu}_C H^{\nu\rho} \nabla_{\nu} T^C_{\mu\rho} \\
    &- \frac{3}{2} T^C_{aC} T^D_{aD} - \inv{8} \eta^{CD} \eta^{EF} f_{abCE} f_{abDF} \bigg)+ O(c^{-6}) \ .
\end{align}
To summarise, upon taking the $c\to\infty$ limit the Einstein equation becomes the three sets of equations
\begin{subequations}
\begin{align} \nonumber
    0 &=  e^{\mu}_a e^{\nu}_b R^{(0)}_{\mn} - \eta_{AB} T^A_{(a|c|} \Tilde{g}^B_{b)c} + \inv{2} T^A_{aA} T^B_{bB} - 3 \eta^{AB} \eta^{CD} f_{acAC} f_{bcBD} - \inv{4}\delta_{ab} \bigg( 2 H^{\mn} R^{(0)}_{\mn} \\ 
    &\qquad - \eta_{AB} T^A_{cd} \Tilde{g}^B_{cd} - T^A_{cA} T^B_{cB} + 2 \tau^{\mu}_A H^{\nu\rho} \nabla_{\nu} T^A_{\mu\rho} + 3 \eta^{AB} \eta^{CD} f_{cdAC} f_{cdBD} \bigg) \ , \\ \nonumber
    0 &=  \tau^{\mu}_{(A} \eta_{B)C} H^{\nu\rho} \nabla_{\nu} T^C_{\mu\rho} - T^D_{aD} T^C_{a(A} \eta_{B)C} + \inv{2} \brac{\eta_{AC} \eta_{BD} \eta^{EF} T^C_{aE} T^D_{aF} - \eta_{CD} T^C_{aA} T^D_{aB} }  \\ \nonumber
    &\qquad+ \inv{2} \eta_{(A|C|} \eta_{B)D} T^C_{ab} \Tilde{g}^D_{ab} -\inv{4} \eta^{CD} f_{ACab} f_{BDab} - \inv{2} \eta_{AB} \bigg(
    H^{\mn} R^{(0)}_{\mn} \\
    &\qquad + 2 \tau^{\mu}_C H^{\nu\rho} \nabla_{\nu} T^C_{\mu\rho} - \frac{3}{2} T^C_{aC} T^D_{aD} - \inv{8} \eta^{CD} \eta^{EF} f_{abCE} f_{abDF} \bigg) \ , \\
    0 &= \eta_{AB} e^{\mu}_a H^{\rho\sigma} \nabla_{\rho} T^B_{\mu\sigma} + \inv{6} \epsilon_{abcde} T^B_{bA} f_{cdeB} - \inv{2} \eta^{BC} f_{abcB} f_{AbcC}  \ ,
\end{align}
\end{subequations}
at leading order.

However, this is not the entire story. Taking the trace of the totally transverse and totally longitudinal equations, we find that the second is twice the first, as predicted by \eqref{eq: eom trace relation}, and so we lose an equation of motion at this order in $c$. We must therefore go to subleading order in $c$ for this projection. The easiest way to do this is to work directly with the projection of the Einstein equation. We find the Einstein and stress tensors become
\begin{subequations}
\begin{align}
    \brac{2 H^{\mn} - c^{-6} \cal{T}^{\mn}} G_{\mn} &= - 3 c^{-6} \cal{T}^{\mn} R_{\mn} \ , \\
    \brac{2 H^{\mn} - c^{-6} \cal{T}^{\mn}} \Theta_{\mn} &= -\frac{3}{2} \brac{c^{-6} \cal{T}^{\mn} - H^{\mn}} \theta_{\mn} \ ,
\end{align}
\end{subequations}
so this component of the Einstein equation is
\begin{equation}
    2 \cal{T}^{\mn} R_{\mn} = \cal{T}^{\mn} \theta_{\mn} - c^6 H^{\mn} \theta_{\mn} \ .
\end{equation}
Expanding this gives
\begin{align} \nonumber
    0 = \ & \frac{c^{18}}{12} f_{abcd} f_{abcd} + c^{12} \bigg[ 
    \inv{2} \eta_{AB} T^A_{ab} T^B_{ab} + \inv{6} \eta^{AB} f_{abcA} f_{abcB} \bigg] \\ \nonumber
    &+ c^6 \bigg[ 
    2 \tau^{\mu}_A H^{\nu\rho} \nabla_{\nu} T^A_{\mu\rho} - 2 T^A_{aA} T^B_{aB} \bigg] + 2 \cal{T}^{\mn} R^{(0)}_{\mn} \\
    &- \inv{6} \eta^{AB} \eta^{CD} \eta^{EF} f_{aACE} f_{aBDF} + O(c^{-6}) \ .
\end{align}
However, using identical calculations to those above we have
\begin{equation}
    \frac{c^{18}}{12} f_{abcd} f_{abcd} = 2 c^6 T^A_{aA} T^B_{aB} + 4 T^A_{aA} g_a + O(c^{-6}) \ ,
\end{equation}
where the first term cancels against the second $O(c^6)$ term, and
\begin{align} \nonumber
    \frac{c^{12}}{2} \brac{\eta_{AB} T^A_{ab} T^B_{ab} + \inv{3} \eta^{AB} f_{abcA} f_{abcB}} &= \frac{c^{12}}{6} \epsilon_{abcde} T^A_{ab} f_{cdeA} + \inv{2} \eta_{AB} \Tilde{g}^A_{ab} \Tilde{g}^B_{ab} \\ \nonumber
    &= \frac{c^{12}}{12} \nabla_{\mu} \brac{\epsilon_{abcde} e^{\mu}_a f_{bcde}} + \inv{2} \eta_{AB} \Tilde{g}^A_{ab} \Tilde{g}^B_{ab} \\
    &= - 2 c^6 \tau^{\mu}_A H^{\nu\rho} \nabla_{\nu} T^A_{\mu\rho} + 2 \nabla_{\mu} g^{\mu} + \inv{2} \eta_{AB} \Tilde{g}^A_{ab} \Tilde{g}^B_{ab} \ ,
\end{align}
with the first term now cancelling against the other $O(c^6)$ term. Taking $c\to\infty$, we find the equation of motion
\begin{equation}\label{eq: Poisson}
    \nabla_{\mu} g^{\mu} + 2 T^A_{aA} g_a = - \cal{T}^{\mn} R^{(0)}_{\mn} + \inv{12} \eta^{AB} \eta^{CD} \eta^{EF} f_{aACE} f_{aBDF} - \inv{4} \eta_{AB} \Tilde{g}^A_{ab} \Tilde{g}^B_{ab} \ .
\end{equation}
This is an equation for the dilatation-covariant divergence of the gradient of the totally longitudinal component of $c_6$, and therefore forms the Poisson equation for our system. It can be checked that both sides of the equation transform covariantly under the local dilatation transformation \eqref{eq: local dilatation symmetry}, and so this symmetry of the action extends to a symmetry of the full set of equations of motion.

\section{Properties of the M5-Brane Limit} \label{sect: properties of the limit}

We have constructed a novel limit of the Bosonic sector of eleven-dimensional supergravity in two seemingly different ways. In order to reconcile the approaches, we should clarify the physical role of the Hubbard-Stratonovich fields $\lambda_{abcd}$ and $\Lambda^A_{ab}$. At finite $c$, the equations of motion for the auxiliary fields are
\begin{subequations}
\begin{align}
    \lambda_{abcd} &= - \frac{c^{12}}{24} \brac{f_{abcd} - c^{-6} \Tilde{T}_{abcd}} \ , \\
    \Lambda^A_{ab} &= - \frac{c^6}{2} \brac{T^A_{ab} - \inv{3!} \epsilon_{abcde} f^{cdeA} } \ ,
\end{align}
\end{subequations}
and so a comparison with the duality relations \eqref{eq: duality constraint 1} and \eqref{eq: duality constraint 2} allow us to make the identifications
\begin{subequations} \label{eq: Lagrange multiplier to 7-form components}
\begin{align}
    \lambda_{abcd} &=  - \inv{24} \epsilon_{abcde} g_e \ , \\
    \Lambda^A_{ab} &= - \inv{240} \epsilon^{ABCDEF} g_{abBCDEF} \ ,
\end{align}
\end{subequations}
between the Hubbard-Stratonovich fields and two projections of the 7-form field strength that enter into the equations of motion. In fact, this identification justifies the odd-looking transformations \eqref{eq: boost transformations} of the Lagrange multiplier fields under five-brane boosts, as a calculation shows that they both arise from the transformation of $g_7$ required to keep the relativistic 7-form field strength invariant.

Another point of interest is the role of M5-branes in the theory. It was found in \cite{Avila:2023aey, Blair:2024aqz} that attempting to take non-relativistic limits aligned with gravitating strings and D-branes leads to an infinite back-reaction, taking us out of the non-relativistic regime\footnote{This point was discussed further in \cite{Harmark:2025ikv, Guijosa:2025mwh}.}. The same reasoning applies in our case to relativistic M5-brane solutions, as can be seen by the following argument analogous to the one in \cite{Blair:2024aqz}. Suppose we start from the relativistic M5-brane solution and scale our vielbeins as in \eqref{eq: M5-brane limit field redefinitions} to get
\begin{subequations} \label{eq: m5 brane scaled solution}
\begin{align}
    ds^2 &= c^2 H^{-1/3} \eta_{AB} dt^A dt^B + c^{-4} H^{2/3} dx^a dx^a \ , \\
    C_6 &= c^6 \brac{H^{-1}-1 } dt^0 \wedge ... \wedge dt^5 \ , \\
    H &= 1 + \frac{c^{\alpha} R^3}{r^3} \ .
\end{align}
\end{subequations}
The $c^{\alpha}$ factor in $H$ is determined by requiring that \eqref{eq: m5 brane scaled solution} is still a solution of the equations of motion with the same source term after the scaling. This means that the number of M5-branes
\begin{equation}
    N_{M5} = \kappa_1 \int_{S^4_{\infty}} F_4 \ ,
\end{equation}
where $\kappa_1$ is an irrelevant proportionality factor, does not scale with $c$. Using \eqref{eq: m5 brane scaled solution} we find
\begin{equation}
    N_{M5} = - \kappa_1 \, c^{-6} \lim_{r\to\infty} \int_{S^4} d\Omega_4 \, r^4 \partial_r H \ ,
\end{equation}
and so we must take $\alpha=6$. Upon taking $c\to\infty$ we see that the factors of $c$ in \eqref{eq: m5 brane scaled solution} cancel, leaving us with the near-horizon limit of the relativistic solution.

At first glance it appears that we lose the ability to describe M5-branes whose worldvolumes are aligned with the limit. One way in which this manifests itself is through the local dilatation symmetry \eqref{eq: local dilatation symmetry}; an M5-brane-like geometry
\begin{subequations}
\begin{align}
    \tau^A &= H^{-1/6} dt^A \ , \\
    e^a &= H^{1/3} dx^a \ ,
\end{align}
\end{subequations}
is equivalent to flat spacetime, ruling out back-reaction of M5-branes on the geometry. However, we can still consider light M5-brane states that do not back-react on the geometry. In fact, by coupling the divergent $(p+1)$-form field to an aligned $p$-brane we remove its rest energy, making this a natural starting point. Let us consider the ansatz
\begin{subequations}
\begin{align}
    \tau^A &= dt^A \ , \\
    e^a &= dx^a \ , \\
    c_3 &= 0 \ , \\
    c_6 &= \phi(r) dt^0 \wedge ... \wedge dt^5 \ ,
\end{align}
\end{subequations}
where $r^2 = x^a x^a$ is the radial coordinate in the transverse directions. The only non-trivial equation of motion is then the Poisson equation, which becomes the transverse Laplace equation for $\phi$; if we include M5-brane sources at the origin this is solved by
\begin{equation}
    g_a = \frac{C x^a}{r^5} \ ,
\end{equation}
and we may then interpret
\begin{equation}
    \frac{8\pi^2 C}{3} = \int_{S^4_{\infty}} dS_a \, g_a
\end{equation}
as the number of M5-branes. Using \eqref{eq: Lagrange multiplier to 7-form components}, we observe that from the Lagrangian perspective the number of M5-branes is given by the flux of the Lagrange multiplier $\lambda_{abcd}$ through the 4-sphere at infinity. Furthermore, as $g_7$ is the field strength of a 6-form gauge field we expect $\lambda$ inherits its flux quantisation condition.

We can also consider the fate of relativistic branes whose worldvolumes are not aligned with the limit. Let us first consider the rescaling of an M5-brane smeared along two directions longitudinal to the limit, which we take to be a torus. The relativistic solution is then
\begin{subequations} \label{eq: transverse M5 brane scaled solution}
\begin{align} \nonumber
    ds^2 &= c^2 \brac{H^{-1/3} \eta_{AB} dt^A dt^B + H^{2/3} d\sigma^I d\sigma^I} \\
    & \qquad + c^{-4} \brac{H^{-1/3} dx^a dx^a + H^{2/3} dY^M dY^M} \ , \\
    C_6 &= \brac{H^{-1} - 1} dt^0 \wedge ... \wedge dt^3 \wedge dx^1 \wedge dx^2 \ , \\
    H &= 1 + \frac{c^{\beta} R}{r} \ ,
\end{align}
\end{subequations}
where $t^A$ are coordinates for $\bb{R}^{1,3}$, $\sigma^I$ coordinates for $T^2$, $x^a$ coordinates for $\bb{R}^2$, and $Y^M$ coordinates for $\bb{R}^3$ with $r^2 = Y^M Y^M$. The divergent part of the 6-form field required by the M5-brane limit is flat for this geometry, and so can be ignored. The exponent $\beta$ is again fixed by requiring that the number of M5-branes, now given by
\begin{equation}
    N_{M5}' = \kappa_2 \int_{S_{\infty}^2 \times T^2} F_4 \ ,
\end{equation}
is finite. From \eqref{eq: transverse M5 brane scaled solution} we see that this is
\begin{equation}
    N_{M5}' = - \kappa_2 \lim_{r\to\infty} \int r^2 \partial_r H \, \epsilon_{S^2} \wedge \epsilon_{T^2} \ ,
\end{equation}
which does not scale with $c$; we must therefore take $\beta= 0$, meaning the back-reacting brane survives the M5-brane limit. 

Similarly, we can consider the scaled smeared M2-brane solution
\begin{subequations} \label{eq: scaled M2-brane solution}
\begin{align} \nonumber
    ds^2 &= c^2 \brac{H^{-2/3} \eta_{AB} dt^A dt^B + H^{1/3} d\sigma^I d\sigma^I} \\
    & \qquad + c^{-4} \brac{H^{-2/3} dx^2 + H^{1/3} dY^M dY^M} \ , \\
    C_3 &= \brac{H^{-1} - 1} dt^0 \wedge dt^1 \wedge dx \ , \\
    H &= 1 + \frac{c^{\gamma} R^2}{r^2} \ ,
\end{align}
\end{subequations}
where $t^A$ are coordinates for $\bb{R}^{1,1}$, $\sigma^I$ coordinates for $T^4$, and $Y^M$ coordinates for $\bb{R}^4$ with $r^2 = Y^M Y^M$. As above, the divergent contribution to the 6-form field is flat for this background and can again be ignored. The M2-branes are smeared over the $T^4$ factor, and the total number of them is given by
\begin{equation}
    N_{M2} = \kappa_3 \int_{S_{\infty}^3 \times T^4} G_7 \ .
\end{equation}
From the solution \eqref{eq: scaled M2-brane solution} we see that
\begin{equation}
    N_{M2} = - \kappa_3 \lim_{r\to\infty} \int_{S^3 \times T^4} r^3 \partial_r H \, \epsilon_{S^3 } \wedge \epsilon_{T^4} \ ,
\end{equation}
and so again we must take $\gamma=0$, with the smeared M2-brane solution surviving the M5-brane limit. The two surviving back-reacting brane solutions we have found are those that arise from quarter-BPS intersecting brane solutions of the relativistic theory, with one of the branes taking on the role of the M5-brane limit. It would be interesting to see if whether these are the only such back-reacting brane solutions in the M5-brane limit, or if other solutions preserving less supersymmetry can be constructed.

Finally, we note that the innocuous looking ansatz 
\begin{subequations}
\begin{align}
    \tau^A &= \tau_{\alpha}^A(t) dt^{\alpha} \ , \\
    e^a &= dx^a \ , \\
    c_3 &= c_6 = 0 \ ,
\end{align}
\end{subequations}
is a solution of the equations of motion provided that the Ricci tensor formed from the Levi-Civita connection of $\cal{T}_{\mn} = \eta_{AB} \tau^A_{\mu} \tau^B_{\nu}$ is traceless,
\begin{equation}
    \cal{T}^{\mn} R_{\mn}(\cal{T}) = 0 \ .
\end{equation}
This includes all solutions of the six-dimensional vacuum Einstein equations, such as the six-dimensional Schwarzschild spacetime \cite{Myers:1986un}
\begin{subequations}
\begin{align}
    \cal{T}_{\alpha\beta} dt^{\alpha} dt^{\beta} &= - f dt^2 + f^{-1} dr^2 + r^2 d\Sigma_{4}^2 \ , \\
    f &= 1 - \frac{C}{r^3} \ ,
\end{align}
\end{subequations}
or gravitational wave solutions, amongst many others.

\section{Conclusion} \label{sect: conclusion}

In this paper we constructed a novel limit of eleven-dimensional supergravity that we associate with a heavy stack of M5-branes. The resulting action is given in \eqref{eq: M5 limit of action} and it is invariant under  Galilean boosts and a local scale symmetry. However we observed that, to completely describe the dynamics, we also need to give the Poisson equation \eqref{eq: Poisson} which is derived by looking at subleading order of  the original  equations of motion and which supplements the equations of motion that  arise from the action. We also gave an interpretation of the Minkowski space solution as describing $N_{M5}$ M5-branes where $N_{M5}$ is determined by the flux of  $\lambda_4$ through the transverse sphere at infinity.

Let us now add some comments and directions for future work.  The consistency of the M5-brane limit as discussed here does not depend on the presence of the Chern-Simons term in the eleven-dimensional supergravity action; it simply goes along for the ride in the action. We note, however, that in the absence of the Chern-Simons term the self-duality constraint $f_{abcd}^{(-)}=0$ of the M2-brane limit would be replaced by the stronger constraint $f_{abcd}=0$. Furthermore, from \eqref{eq: C6M2} we see that the $C_3\wedge dC_3$ term in the duality relation (which arises from the Chern-Simons term in the action) leads to a divergent contribution to $C_6$ and hence $G_7$. The presence of the Chern-Simons term in the relativistic theory is essential for the theory to be supersymmetric, and one may wonder if it serves a similar role in the M5-brane limit. Determining whether or not supersymmetry is preserved appears to be a very difficult problem for the case of the  M2-brane limit \cite{Bergshoeff:2024nin}; however, one might hope that the argument presented in \cite{Lambert:2019nti} (but only valid for rigid supersymmetry) could be adapted to the M5-brane limit considered here. Aside from the requirement of supersymmetry, it has previously been argued that the Chern-Simons term is also required for the M2-brane and M5-brane solutions to satisfy a lower bound on the tension/charge ratio \cite{Gibbons:1994vm}.  It would be interesting to see whether a similar argument for the non-relativistic M2-brane and M5-brane solutions can be given.

Another open question is the extent to which the dualities of the relativistic theory extend to the non-relativistic limits; can we still view the non-relativistic M5-brane as the magnetic dual to the non-relativistic M2-brane? What are the U-duality symmetries that one finds after toroidal compactification? Similarly, we could ask if a version of the AdS/CFT correspondence survives in the M5-brane limit of M-theory. In \cite{Lambert:2024yjk} the M5-brane limit of the worldvolume CFT of transverse M2 was studied. It was found that the limit localises it onto near-BPS states describing M5-branes, aligned with the limit, on which the M2-brane's worldvolume ends. The orientation of the worldvolumes and limit are exactly those of the M2 solution considered in section \ref{sect: properties of the limit}, and so by going to the near-horizon regime it appears that we are taking the same limit on both sides of the holographic duality. It is therefore natural to conjecture that the duality is preserved by the limit. This proposal satisfies the obvious zeroth order tests. For instance in \cite{Lambert:2024ncn} it was  shown that the Bosonic symmetries preserved by the limit match on both sides, but assuming local scale symmetry of the M5-brane limit of eleven-dimensional supergravity which we have now established here. It is of interest to examine this duality and the M5-brane theory further.

\section*{Acknowledgments}

E.B. would like to thank Luca Romano for useful discussions. N.L. and J.S. would like to thank the University of Groningen for its hospitality. N.L. is supported in part  by the STFC consolidated grant ST/X000753/1. J.S. is supported by the STFC studentship ST/W507556/1. This work is supported by the Croatian Science Foundation project IP-2022-10-5980 “Non-relativistic supergravity and applications''. 

\appendix

\section{Non-Relativistic Expansion of the Curvature} \label{sect: curvature expansion}

In this appendix we elaborate on the non-relativistic expansion of the curvature utilised in sections \ref{sect: M5-brane limit of action} and \ref{sect: einstein equation limit}. The relativistic Ricci tensor is defined by
\begin{equation} \label{eq: ricci tensor def}
    R_{\mn} = \partial_{\rho} \hat{\Gamma}^{\rho}_{\nu\mu} - \partial_{\nu} \hat{\Gamma}^{\rho}_{\rho\mu} + \hat{\Gamma}^{\sigma}_{\nu\mu} \hat{\Gamma}^{\rho}_{\rho\sigma} - \hat{\Gamma}^{\sigma}_{\rho\mu} \hat{\Gamma}^{\rho}_{\nu\sigma} \ ,
\end{equation}
where $\hat{\Gamma}$ is the Levi-Civita connection of $g_{\mn}$. After performing the rescaling \eqref{eq: M5-brane limit field redefinitions} of the vielbeins, the connection has the expansion
\begin{equation}
    \hat{\Gamma}^{\rho}_{\mn} = c^6 A^{\rho}_{\mn} + B^{\rho}_{\mn} + c^{-6} C^{\rho}_{\mn} \ ,
\end{equation}
with
\begin{subequations}
\begin{align}
    A^{\rho}_{\mn} &= \eta_{AB} H^{\rho\sigma} \tau_{(\mu}^A T_{\nu)\sigma}^B \ , \\
    B^{\rho}_{\mn} &= \Gamma^{\rho}_{\mn} + X^{\rho}_{\mn} \ , \\
    C^{\rho}_{\mn} &= \cal{T}^{\rho\sigma}    e^a_{(\mu} E^a_{\nu)\sigma}  \ .
\end{align}
\end{subequations}
We have split the $O(c^0)$ piece into the non-relativistic connection $\Gamma$ defined in equation \eqref{eq: NR connection} and $X$, which we take to be
\begin{equation}
     X^{\rho}_{\mn} = \eta_{AB} \cal{T}^{\rho\sigma} \tau^A_{(\mu} T_{\nu)\sigma}^B - \inv{2} \tau_A^{\rho} T_{\mn}^A \ .
\end{equation}
Inserting this expansion into \eqref{eq: ricci tensor def} and using
\begin{equation}
    A^{\rho}_{\rho\mu} = 0\ ,
\end{equation}
gives
\begin{align} \nonumber
    R_{\mn} = & \, - c^{12} A^{\sigma}_{\mu\rho} A^{\rho}_{\nu\sigma} + c^6 \Big(
    \nabla_{\rho} A^{\rho}_{\mn} - \tau^{\rho}_A T^A_{\nu\sigma} A^{\sigma}_{\mu\rho} + A^{\sigma}_{\mn} X^{\rho}_{\rho\sigma} - A^{\rho}_{\mu\sigma} X^{\sigma}_{\nu\rho} \\ \nonumber
    &- A^{\rho}_{\sigma\nu} X^{\sigma}_{\rho\mu} \Big) + \cal{R}_{\mn} + \nabla_{\rho} X^{\rho}_{\nu\mu} - \nabla_{\nu} X^{\rho}_{\rho\mu} - \tau^{\rho}_A T^A_{\nu\sigma} 
    X^{\sigma}_{\rho\mu}
    + X^{\sigma}_{\nu\mu} X^{\rho}_{\rho\sigma} \\
    &- X^{\sigma}_{\rho\mu} X^{\rho}_{\nu\sigma} + A^{\sigma}_{\mn} C^{\rho}_{\rho\sigma} - A^{\sigma}_{\mu\rho} C^{\rho}_{\nu\sigma} - A^{\sigma}_{\nu\rho} C^{\rho}_{\mu\sigma} + O(c^{-6}) \ ,
\end{align}
where $\cal{R}_{\mn}$ is the Ricci tensor of $\Gamma$ and is defined using the same conventions as \eqref{eq: ricci tensor def}, and as above $\nabla$ is the covariant derivative associated with $\Gamma$. However, using the identity
\begin{equation}
    2 X^{\rho}_{[\mu\nu]} = -\tau^{\rho}_A T^A_{\mn} \ ,
\end{equation}
we see this simplifies to
\begin{align} \nonumber
    R_{\mn} = & \, - c^{12} A^{\sigma}_{\mu\rho} A^{\rho}_{\nu\sigma} + c^6 \Big(
    \nabla_{\rho} A^{\rho}_{\mn} + A^{\sigma}_{\mn} X^{\rho}_{\rho\sigma} - A^{\rho}_{\mu\sigma} X^{\sigma}_{\rho\nu} - A^{\rho}_{\sigma\nu} X^{\sigma}_{\rho\mu} \Big) \\ \nonumber
    &+ \cal{R}_{\mn} + \nabla_{\rho} X^{\rho}_{\nu\mu} - \nabla_{\nu} X^{\rho}_{\rho\mu} 
    + X^{\sigma}_{\nu\mu} X^{\rho}_{\rho\sigma} - X^{\sigma}_{\rho\mu} X^{\rho}_{\sigma\nu} + A^{\sigma}_{\mn} C^{\rho}_{\rho\sigma} \\
    &- A^{\sigma}_{\mu\rho} C^{\rho}_{\nu\sigma} - A^{\sigma}_{\nu\rho} C^{\rho}_{\mu\sigma} + O(c^{-6}) \ .
\end{align}
It will be simplest to express this in terms of its projections onto the longitudinal and transverse basis vectors, where we have
\begin{subequations}
\begin{align}
    e^{\mu}_a e^{\nu}_b R_{\mn} &= - \frac{c^6}{2} \eta_{AB} T^A_{ac} T^B_{bc} + e^{\mu}_a e^{\nu}_b R_{\mn}^{(0)} + O(c^{-6}) \ , \\
    e^{\mu}_a \tau^{\nu}_A R_{\mn} &= \frac{c^6}{2} \brac{
    \eta_{AB} e^{\mu}_a H^{\rho\sigma} \nabla_{\rho} T_{\mu\sigma}^B - 2\eta_{B(C}  T^C_{A)b} T^B_{ab} } + e^{\mu}_a \tau^{\nu}_A R_{\mn}^{(0)} + O(c^{-6}) \ , \\ \nonumber
    \tau^{\mu}_A \tau^{\nu}_B R_{\mn}  &=  \, \frac{c^{12}}{4} \eta_{AC} \eta_{BD} T^C_{ab} T^D_{ab} + c^6 \bigg[
    \tau^{\mu}_{(A} \eta_{B)C} H^{\nu\rho} \nabla_{\nu } T_{\mu\rho}^C - T^D_{aD} T^C_{a(A} \eta_{B)C} \\
    &\qquad + \inv{2} \brac{
    \eta_{AC} \eta_{BD} \eta^{EF} T^C_{aE} T^D_{aF} - \eta_{CD} T^C_{aA} T^D_{aB} } \bigg] + O(c^0) \ .
\end{align}
\end{subequations}
From this, we can take the trace to find
\begin{subequations}
\begin{align}
    H^{\mn} R_{\mn} &= - \frac{c^6}{2} \eta_{AB} T^A_{ab} T^B_{ab} + H^{\mn} R_{\mn}^{(0)} + O(c^{-6}) \ , \\
    \cal{T}^{\mn} R_{\mn} &= \frac{c^{12}}{4} \eta_{AB} T^A_{ab} T^B_{ab} - c^6 \bigg[  H^{\mn} \tau_A^{\rho} \nabla_{\mu} T^A_{\nu\rho} + T^A_{aA} T^B_{aB} \bigg] + O(c^0) \ ,
\end{align}
\end{subequations}
To proceed further we must simplify the finite term in the spatial trace. As we have the identities
\begin{subequations}
\begin{align}
    H^{\mn} X_{\mn} &= 0 \ , \\
    H^{\mn} X^{\sigma}_{\rho\mu} X^{\rho}_{\sigma\nu} &= \inv{2} \brac{T^A_{aB} T^B_{aA} + \eta_{AB} \eta^{CD} T^A_{aC} T^B_{aD}} \ , \\
    H^{\mn} A^{\sigma}_{\mu\rho} C^{\rho}_{\nu\sigma} &= 0 \ ,
\end{align}
\end{subequations}
we are left with
\begin{equation}
    H^{\mn} R_{\mn}^{(0)} = H^{\mn} \cal{R}_{\mn} - H^{\mn} \nabla_{\mu} X^{\rho}_{\rho\nu} - \inv{2} \brac{T^A_{aB} T^B_{aA} + \eta_{AB} \eta^{CD} T^A_{aC} T^B_{aD}} \ .
\end{equation}
If we use
\begin{equation}
    X^{\rho}_{\rho\mu} = T^A_{\mn} \tau^{\nu}_A
\end{equation}
we find that the remaining term is just
\begin{equation}
    H^{\mn} \nabla_{\mu} X^{\rho}_{\rho\nu} = H^{\mn} \tau^{\rho}_A \nabla_{\mu} T^A_{\nu\rho} + T^A_{a \nu} e^{\mu}_a \nabla_{\mu} \tau_A^{\nu} \ .
\end{equation}
However, a short calculation gives
\begin{align} \nonumber
    T^A_{a\nu} e^{\mu}_a \nabla_{\mu} \tau^{\nu}_A &=  T^A_{ab} \tau_A^{\mu} \brac{ e^{\nu}_a \partial_{[\mu} e^b_{\nu]} + e^{\nu}_b \partial_{[\mu} e^a_{\nu]} } \\
    &= 0 \ ,
\end{align}
due to the antisymmetry of $T^A$, and so the trace becomes
\begin{equation}
    H^{\mn} R_{\mn}^{(0)} = H^{\mn} \cal{R}_{\mn} - H^{\mn} \tau^{\rho}_A \nabla_{\mu} T^A_{\nu\rho} - \inv{2} \brac{T^A_{aB} T^B_{aA} + \eta_{AB} \eta^{CD} T^A_{aC} T^B_{aD}} \ .
\end{equation}
The Ricci scalar
\begin{equation}
    R = c^4\brac{H^{\mn} + c^{-6} \cal{T}^{\mn}} R_{\mn}\ ,
\end{equation}
therefore has the expansion
\begin{align} \nonumber
    R = \, & - \frac{c^{10}}{4} \eta_{AB} T^A_{ab} T^B_{ab} + c^4 \bigg[ 
    H^{\mn} \cal{R}_{\mn} - 2 H^{\mn} \tau^{\rho}_A \nabla_{\mu} T^A_{\nu\rho}  \\
    &- \inv{2} \brac{T^A_{aB} T^B_{aA} + \eta_{AB} \eta^{CD} T^A_{aC} T^B_{aD}} - T^A_{aA} T^B_{aB} \bigg] + O(c^{-2}) \ .
\end{align}

Finally, we note that as $\Gamma$ has non-vanishing torsion the variation of the non-relativistic Ricci tensor is given by
\begin{equation} \label{eq: Ricci tensor variation}
    \delta \cal{R}_{\mn} = \nabla_{\rho} \delta \Gamma^{\rho}_{\nu\mu} - \nabla_{\nu} \delta \Gamma^{\rho}_{\rho\mu} - \tau^{\rho}_A T^A_{\nu\sigma} \delta \Gamma^{\sigma}_{\rho\mu} \ .
\end{equation}

\printbibliography

\end{document}